\documentclass[floatfix,twocolumn,showpacs,preprintnumbers,amsmath,amssymb,prr,superscriptaddress,longbibliography, nofootinbib]{revtex4-1}
\usepackage{color}
\usepackage[usenames,dvipsnames,svgnames,table]{xcolor}
\usepackage[colorlinks=true,linkcolor=blue,urlcolor=blue,citecolor=blue]{hyperref}
\usepackage{mathtools}
\usepackage{graphicx}
\usepackage{dcolumn}
\usepackage{array}
\usepackage{lipsum}
\usepackage{bm}
\usepackage{subfigure}
\usepackage{amssymb}
\usepackage{multirow}
\usepackage{tabularx}
\usepackage{amsmath}
\usepackage{braket}
\usepackage{csquotes}
\graphicspath{{plots/}}
 \usepackage{lipsum}
\usepackage{mathrsfs}
\usepackage{MnSymbol}

\newcommand{\beq}{\begin{equation}}
\newcommand{\eeq}{\end{equation}}
\newcommand{\bea}{\begin{eqnarray}}
\newcommand{\eea}{\end{eqnarray}}

\renewcommand{\vec}[1]{\mathbf{#1}}

\begin{document}
\title{ 
Excitation signatures of isochorically heated electrons in solids at finite wavenumber explored from first principles
}

\author{Zhandos~A.~Moldabekov}
\email{z.moldabekov@hzdr.de}

\affiliation{Center for Advanced Systems Understanding (CASUS), D-02826 G\"orlitz, Germany}
\affiliation{Helmholtz-Zentrum Dresden-Rossendorf (HZDR), D-01328 Dresden, Germany}

\author{Thomas D. Gawne}

\affiliation{Center for Advanced Systems Understanding (CASUS), D-02826 G\"orlitz, Germany}
\affiliation{Helmholtz-Zentrum Dresden-Rossendorf (HZDR), D-01328 Dresden, Germany}

\author{Sebastian Schwalbe}

\affiliation{Center for Advanced Systems Understanding (CASUS), D-02826 G\"orlitz, Germany}
\affiliation{Helmholtz-Zentrum Dresden-Rossendorf (HZDR), D-01328 Dresden, Germany}

\author{Thomas~R.~Preston}
\affiliation{European XFEL, D-22869 Schenefeld, Germany}

\author{Jan Vorberger}
\affiliation{Helmholtz-Zentrum Dresden-Rossendorf (HZDR), D-01328 Dresden, Germany}

\author{Tobias Dornheim}

\affiliation{Center for Advanced Systems Understanding (CASUS), D-02826 G\"orlitz, Germany}
\affiliation{Helmholtz-Zentrum Dresden-Rossendorf (HZDR), D-01328 Dresden, Germany}

\begin{abstract}
Ultrafast heating of solids with modern X-ray free electron lasers (XFELs) leads to a unique set of conditions that is characterized by the simultaneous presence of heated electrons in a cold ionic lattice.
In this work, we analyze the effect of electronic heating on the dynamic structure factor (DSF) in bulk Aluminium (Al) with a face-centered cubic lattice and in silicon (Si) with a crystal diamond structure using first-principles linear-response time-dependent density functional theory simulations.
We find a thermally induced red shift of the collective plasmon excitation in both materials.
In addition, we show that the heating of the electrons in Al can lead to the formation of a double-plasmon peak due to the extension of the Landau damping region to smaller wavenumbers. Finally, we demonstrate that thermal effects generate a measurable and distinct signature (peak-valley structure) in the DSF of Si at small frequencies.  Our simulations indicate that there are a variety of new features in the spectrum of X-ray-driven solids, specifically at finite momentum transfer, which can be probed in upcoming X-ray Thomson scattering (XRTS) experiments at various XFEL facilities.
\end{abstract}


\maketitle

\section{Introduction}

First-principles simulation methods 
are well established for the theory of solids at ambient conditions where the electrons are in their respective ground state,
enabling the prediction of the electronic and structural characteristics of a variety of materials \cite{Lithium_Ammonia, Alkali_Metals, Si_Weissker}. 
In contrast, the properties of the hot, correlated electrons generated by laser heating of solids remain 
substantially less explored.
Such systems are located in the \emph{warm dense matter} (WDM) regime~\cite{wdm_book,new_POP,review}, which also naturally occurs in a host of astrophysical objects~\cite{drake2018high,Benuzzi_Mounaix_2014,becker}, and as a transient state in technological applications such as laser fusion~\cite{hu_ICF,Betti2016}.
In WDM, the thermal energy of electrons is comparable with the Fermi energy and, at the same time, inter-particle correlations  
remain strong~\cite{Ott2018}. This is different from ordinary plasma states where thermal and quantum kinetic effects dominate over correlation effects.
Recent  breakthroughs in inertial confinement fusion research~\cite{Betti2023}, as well as 
advances in hot-electron chemistry~\cite{Brongersma2015} and material science~\cite{Kraus2016,Lazicki2021}  have rendered the accurate understanding of WDM a highly active frontier. 

Although the typical parameters (e.g.~time scales) in inertial fusion applications differ from the WDM conditions that we study here, 
an accurate description of WDM is needed to understand the full compression path of a fuel capsule~\cite{hu_ICF}.
In practice, however, the theoretical understanding of WDM states remains limited due to the 
complex interplay between exchange-correlation effects, quantum degeneracy, and thermal excitations. Testing existing theories and approximations requires reliable experimental data with well enough defined system parameters. In WDM experiments, two types of systems can be distinguished depending on the time scale of measurements: those with fully equilibrated electrons and ions \cite{DOPPNER2009182, Dornheim_T_2022}, and those in non-equilibrium \cite{Hoidn_PRB_2018, Kieffer_2021,Vorberger_PRX_2023}. 
A particularly interesting sub-set of non-equilibrium is given by effective two-temperature systems where both the electrons and ions are approximately in a local state of equilibrium, but with two different temperatures~\cite{ernstorfer,Mo_2018}. 
Such experiments can be performed at X-ray free electron laser (XFEL) facilities such as the European XFEL \cite{Zastrau_2021,Tschentscher_2017} in Germany and LCLS at SLAC \cite{Ding_2009, Fletcher2015} in the USA, and rely on high brightness sources with narrow bandwidth.
At these facilities, the ultrafast isochoric heating of solids with XFELs can create a transient state of matter that features hot electrons within the still cold crystal structure of the ions \cite{Descamps_sciadv, Descamps2020}.
Since the system density and lattice parameters are well-defined, such experiments can provide valuable data for validating state-of-the-art theories and simulation methods for electronic structures in the WDM regime.

For X-ray-driven heating, the incident X-rays cannot directly interact with the ions. Ion heating is therefore primarily driven by electron-phonon interactions which occur on the timescales of several picoseconds~\cite{Ng1995,Matthieu2011,White2014}. As typical free electron laser (FEL) pulses have sub-100~fs duration, the ions remain cold during the FEL pulse \cite{Descamps_sciadv}.
Evidently, the extreme conditions and short timescales render the rigorous diagnostics of these exotic states challenging. In this work, we focus on the X-ray Thomson scattering (XRTS) technique~\cite{siegfried_review,sheffield2010plasma}, which can meet these demands by effectively measuring the electronic dynamic structure factor (DSF) $S_{ee}(\mathbf{q},\omega)$. Indeed, XRTS has emerged as a standard diagnostic method for WDM experiments, as it is, in principle, capable of giving one access to important system parameters such as the temperature, density, and ionization~\cite{kraus_xrts,Tilo_Nature_2023,Gregori_PRE_2003}.

From a practical perspective, the properties of hot electrons in the crystal structure of the ions are relevant for hot-electron chemistry \cite{Brongersma2015, Lin_PRB_2008}.
Moreover, the very controlled conditions and potentially unrivaled diagnostic capabilities (such as high brilliance~\cite{Fletcher2015}, high resolution~\cite{Wollenweber_RSI,Descamps2020}, and high repetition rate~\cite{Tschentscher_2017}) at XFEL facilities makes such experiments a much needed rigorous benchmark for dynamical simulation methods for the description of WDM systems.


Often, the uniform electron gas (UEG) model~\cite{loos,review} is used to provide a qualitative picture of the electrons in metals~\cite{mahan1990many,quantum_theory}. 
For example, in bulk Aluminium (Al) with a face-centered cubic (fcc) lattice, the dependence of the plasmon dispersion on wavenumber has a similar functional form to that of the UEG \cite{Lee_PRB}.
However, the plasma frequency of valence electrons in fcc Al $\omega_p\simeq 15~{\rm eV}$ differs from that of the UEG $\omega_p\simeq 16~{\rm eV}$ at the same density. 
The free electron model becomes quantitatively accurate at high temperatures in plasmas as the electron-ion coupling diminishes. Therefore, one can expect the response properties of electrons in metals to approach that of a free electron gas with increasing temperature. 
In the present work, we compute the DSF of isochorically heated electrons using first-principles linear-response time-dependent density-functional theory (LR-TDDFT) simulations~\cite{book_Ullrich} to quantify the change in the XRTS signal with increasing temperature. Specifically, we consider the crystal structure of fcc Al and crystal diamond silicon (Si).

Several interesting features emerge due to thermal excitations in both materials. Surprisingly, we observe that the DSF of Al remains quantitatively different from the UEG model even at relatively high temperatures in the range of up to $T\sim10~{\rm eV}$. 
Furthermore, we report several interesting observations emerging due to thermal excitations:
(i) Our LR-TDDFT results show a reduction of the plasma frequency in Si and Al at small wavenumbers with increasing temperatures up to a few eV;
(ii) We observe the formation of a double-plasmon peak in Al caused by the shift of the Landau damping region to smaller wavenumbers;
(iii) A thermal-excitation induced signature appears in the DSF of Si at small energies, which is absent in the ground state. 
Finally, we discuss the feasibility of detecting these signatures in experiments at XFEL facilities such as the European XFEL (Germany), LCLS  (USA), and SACLA (Japan).
Our study can serve as a guide 
for the planning of upcoming experiments and clearly indicate the required specifications to resolve the predicted effects with high confidence.


The paper is organized as follows: In Sec. \ref{s:methods}, we describe the simulation method and details. After that the results are presented in Sec. \ref{s:results}.
We discuss the feasibility of measuring the features induced by thermal excitations in Sec. \ref{s:exp}.
We conclude the paper by summarizing our findings and providing an outlook. 

\section{Methods and Simulation Details}\label{s:methods}

\begin{figure}
    \centering
    \includegraphics[width=0.48\textwidth,keepaspectratio]{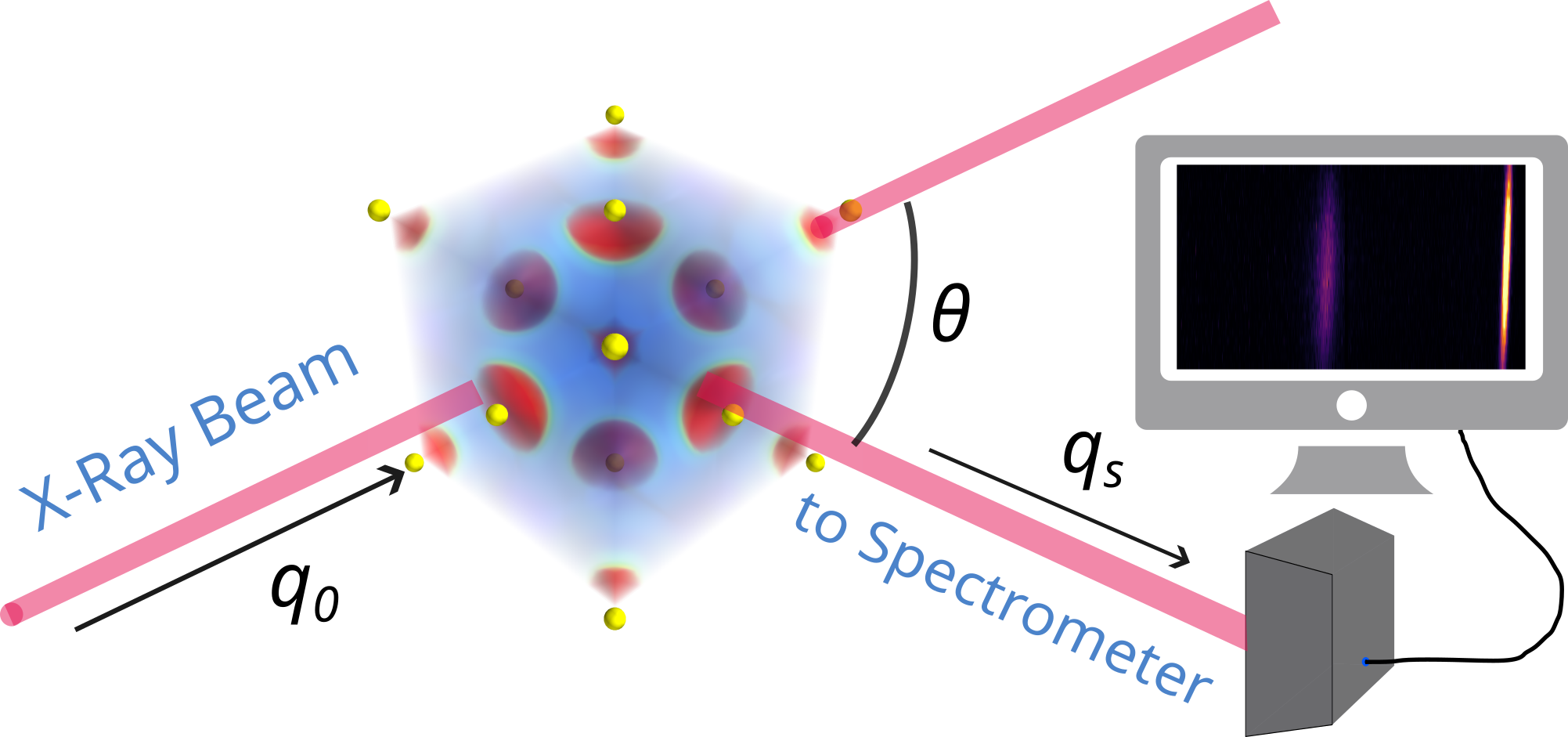}
    \caption{Illustration of an XRTS setup in experiments where the XFEL is used to probe a target. The XFEL beam with frequency (energy) $\hbar\omega_0$ and wavenumber $\mathbf{q_0}$ is scattered at an angle $\theta$ and detected by a spectrometer providing spectrally resolved XRTS data. The frequency $\omega$ and the wavenumber $\vec q$ (i.e., momentum transfer ) in the DSF in Eq. (\ref{eq:double_dif_cross_section})  are defined by energy and momentum conservation: $\hbar\omega=\hbar\omega_0-\hbar\omega_s$ and $\mathbf{q}=\mathbf{q_0}-\mathbf{q_s}$.}
    \label{fig:illus}
\end{figure}

The double-differential cross-section that is measured in an XRTS experiment (as illustrated in Fig. \ref{fig:illus}) is directly proportional to the electronic DSF~\cite{Crowley_2013},
\begin{align}
    \label{eq:double_dif_cross_section}
    \left[\frac{ \partial^{2} \tilde{\sigma} }{ \partial\Omega\partial\omega_{\text{s}} }\right]_{\rm XRTS}
    = &\,
    \frac{\omega_{\text{s}}}{\omega_{0}}
    \left. \frac{ \partial \tilde{\sigma} }{ \partial \Omega } \right|_{\mathrm{T}} \, 
    S_{ee}(\mathbf{q},\omega)
    \,,
\end{align}
where $\left.\partial \tilde{\sigma} / \partial \Omega\right|_{\mathrm{T}}$ is the differential Thomson cross-section with respect to the solid angle, $\omega_s$ is the frequency of the scattered X-rays, and $\omega_0$ is the frequency of the incident XFEL beam.
In practice, an additional difficulty comes from the effects of the employed detector and the finite width of the X-ray laser beam that are described by the combined source-and-instrument function $R(\omega)$. The actually detected XRTS intensity can then be expressed by a convolution of $S_{ee}(\mathbf{q},\omega)$ with $R(\omega)$,
\begin{eqnarray}\label{eq:convolution}
    I(\mathbf{q},\omega) = S_{ee}(\mathbf{q},\omega) \circledast R(\omega)\ .
\end{eqnarray}

The total DSF $S_{ee}(\mathbf{q},\omega)$ can be further decomposed into a quasi-elastic contribution $S_\textnormal{el}(\mathbf{q},\omega)\sim\delta(\omega)$ due to the bound electrons and the screening cloud~\cite{Vorberger_PRE_2015}, and the inelastic contribution $S_\textnormal{inel}(\mathbf{q},\omega)$ that contains transitions between free and bound electrons in any combination~\cite{boehme2023evidence}.
By definition, the quasi-elastic part of the intensity emerges in the limit of small $\omega$ and its shape is defined by $R(\omega)$. For example, at the SASE2 undulator of the European XFEL, a self-seeded XFEL beam with narrow bandwidth $\Omega<1~\rm{eV}$ is routinely achieved ~\cite{Zastrau_2021}, with the shape being nearly Gaussian.  
In this work, we compute the inelastic part of the total DSF using the LR-TDDFT method. In the following, we will simply denote it as $S(\mathbf{q},\omega)$. The quasi-elastic part can be computed if the static structure factors for different pairs of the particle species are known \cite{Vorberger_PRE_2015}, e.g, using KS-DFT based molecular dynamics simulations.

\subsection{LR-TDDFT calculations}

The DSF is computed from the fluctuation-dissipation theorem~\cite{Kubo_1966, quantum_theory},
\begin{equation}
    S(\vec q, \omega)=-\frac{\hbar^2q^2}{4\pi^2e^2n}~\frac{1}{1-e^{-\hbar \omega/k_BT}}~{\rm Im}\left[\epsilon_M^{-1}(\vec q, \omega)\right],
\end{equation}
where $\epsilon_M(\vec q, \omega)$ is the macroscopic dielectric function.

Within LR-TDDFT, the inverse of the macroscopic dielectric function is defined as the diagonal part of the inverse microscopic dielectric matrix $\epsilon_M^{-1}(\vec q, \omega)= \left[\epsilon^{-1}(\vec k, \omega)\right]_{\vec G\vec G}$, where $\vec q=\vec G+ \vec k$ and $\vec G$ is a reciprocal lattice vector \cite{DSF_LR-TDDFT, martin_reining_ceperley_2016}. 

The inverse microscopic dielectric matrix is computed from the microscopic density response function ~\cite{book_Ullrich},
\begin{equation}\label{eq:d_f}
      \varepsilon^{-1}_{\scriptscriptstyle \vec G,\vec G^{\prime}}(\vec k,\omega)=\delta_{\scriptscriptstyle \vec G,\vec G^{\prime}}+\frac{4\pi}{\left|\vec k+\vec G\right|^{2}}  \chi_{\scriptscriptstyle \vec G,\vec G^{\prime}} (\vec k,\omega),
\end{equation}
with  $\vec k$ being restricted to the first Brillouin zone.

The Fourier coefficients $  \chi_{\scriptscriptstyle \vec G,\vec G^{\prime}}(\vec k,\omega)$ of the density response matrix are the solution of a Dyson’s type equation  \cite{Byun_2020, martin_reining_ceperley_2016}:
\begin{equation}\label{eq:Dyson}
\begin{split}
\chi_{\scriptscriptstyle \mathbf G \mathbf G^{\prime}}(\mathbf k, \omega)
&= \chi^0_{\scriptscriptstyle \mathbf G \mathbf G^{\prime}}(\mathbf k, \omega)+ \displaystyle\smashoperator{\sum_{\scriptscriptstyle \mathbf G_1 \mathbf G_2}} \chi^0_{\scriptscriptstyle \mathbf G \mathbf G_1}(\mathbf k, \omega) \big[ v_{\scriptscriptstyle \mathbf G1}(\vec k)\delta_{\scriptscriptstyle \mathbf G_1 \mathbf G_2} \\
&+ K^{\rm xc}_{\scriptscriptstyle \mathbf G_1 \mathbf G_2}(\mathbf k, \omega) \big]\chi_{\scriptscriptstyle \mathbf G_2 \mathbf G^{\prime}}(\mathbf k, \omega),
\end{split}
\end{equation}
where   $\chi^{~0}_{\scriptscriptstyle \vec G,\vec G^{\prime}}(\vec k,\omega)$ is an ideal (non-interacting) density response function computed using Kohn-Sham (KS) orbitals \cite{Hybertsen}, $v_{\scriptscriptstyle \mathbf G1}(\vec k)={4\pi}/{|\mathbf k+\mathbf G_1|^2}$ is the Coulomb potential in reciprocal-space, and $ K^{\rm xc}_{\scriptscriptstyle \vec G_1,\vec G_2}(\vec k, \omega)$ is the exchange-correlation (XC) kernel defined as functional derivative of the exchange-correlation potential \cite{Gross_PRL1985}.

For real materials, the full dynamic XC-kernel is generally unknown. In this situation,
LR-TDDFT with a static (adiabatic) XC kernel   $K^{\rm xc}_{\scriptscriptstyle \vec G_1,\vec G_2}(\vec k, \omega=0)$ in the local density approximation (ALDA) often provides accurate results for the plasmon dispersion that are in agreement with experiments for Al \cite{Quong_PRL, Cazzaniga_2011} and Si \cite{Weissker_PRL, Si_Weissker} at ambient conditions. Moreover, the ALDA approximation was successfully used by  Mo \textit{et al.}~\cite{Mo_2018} to describe the XRTS spectrum of  Al in the WDM regime~\cite{Sperling_2015}.
Therefore, we employ the ALDA in the current work, and
 the presented results can potentially be refined either using more advanced static XC kernels \cite{Moldabekov_JCTC_2023, Moldabekov_PRR_2023, JCP_averaging, Moldabekov_non_empirical_hybrid} or available approximations for a dynamic XC kernel \cite{Byun_2020, PRL_Bootstrap, PRL_Panholzer}.
We note that an alternative approach to LR-TDDFT for computing the DSF within the realm of DFT is the real-time TDDFT (RT-TDDFT) method~\cite{Baczewski_prl_2016}.
If the same XC functional were used in both, RT-TDDFT and LR-TDDFT are formally equivalent for linear response properties ~\cite{book_Ullrich}.

 
\subsection{Simulation parameters}

The LR-TDDFT simulations have been performed using the GPAW code~\cite{GPAW1, GPAW2, LRT_GPAW1, LRT_GPAW2, ase-paper, ase-paper2}, which is a real-space implementation of the projector augmented-wave (PAW) method~\cite{BlochlPAW}. The lattice parameters of fcc Al and crystal diamond Si were set to $a=4.05 ~{\rm \AA}$ and  $a=5.431 ~{\rm \AA}$, respectively (according to experimental observations)~\cite{wyckoff1948crystal}.
The energy cutoff is set to $1000~{\rm eV}$ for both Al and Si. The XC functional in all calculations was set to the ground-state LDA by Perdew and Wang \cite{Perdew_Wang}.
PAW datasets provided by GPAW have been used. 

In the case of Al, for the DSF calculations along the [001] crystallographic direction, we set the $k$-point grid to $40\times40\times55$.
In the DSF simulations along the direction [011], we set the $k$-point grid to $54\times54\times54$. For Al, we consider a temperature range from $0.025~{\rm eV}$ up to $10~\rm{eV}$.
For the considered primitive cell, we used $N_b=100$ KS bands at $T<5 ~\rm{eV}$, $N_b=200$ at $5\leq T< 9~\rm{eV}$, and $N_b=350$ at $T\geq 9~\rm{eV}$.
The local field effect cutoff included in the dielectric function was set to $120~{\rm eV}$. The Lorentzian smearing parameter of the calculations for Al was set to $\eta=0.05~{\rm eV}$.
For Al, we consider wavenumbers in the range from $0.05~{\rm \AA^{-1}}$ up to $1.21~{\rm \AA^{-1}}$.

In the case of Si, for the considered primitive cell, we used  $N_b=100$ for $T=0.025~{\rm eV}$, $T=2~{\rm eV}$, and $T=4~{\rm eV}$.
For the direction [001] we used $\eta=0.05~{\rm eV}$ with $20\times20\times42$ and $40\times40\times40$  $k$-point grids. For the direction [111], the grid was set to $80\times80\times80$  and $\eta=0.1~{\rm eV}$. The local field effect cutoff in the dielectric function was set to $150~{\rm eV}$.
For Si, we consider wavenumbers in the range from $0.0275~{\rm \AA^{-1}}$ up to $1.5~{\rm \AA^{-1}}$.

In all cases, the ions have been treated as frozen in their crystal lattice positions. This approximation is justified by the electron-phonon interaction time being long compared to typical FEL pulse durations, with the entire scattering signal being produced while the pulse is incident on the target. The ions, therefore, remain cold during the scattering process, and the displacement of a cold ion during the scattering time is negligible.

\begin{figure*}[t!]
    \centering
    \includegraphics[width=0.9\textwidth,keepaspectratio]{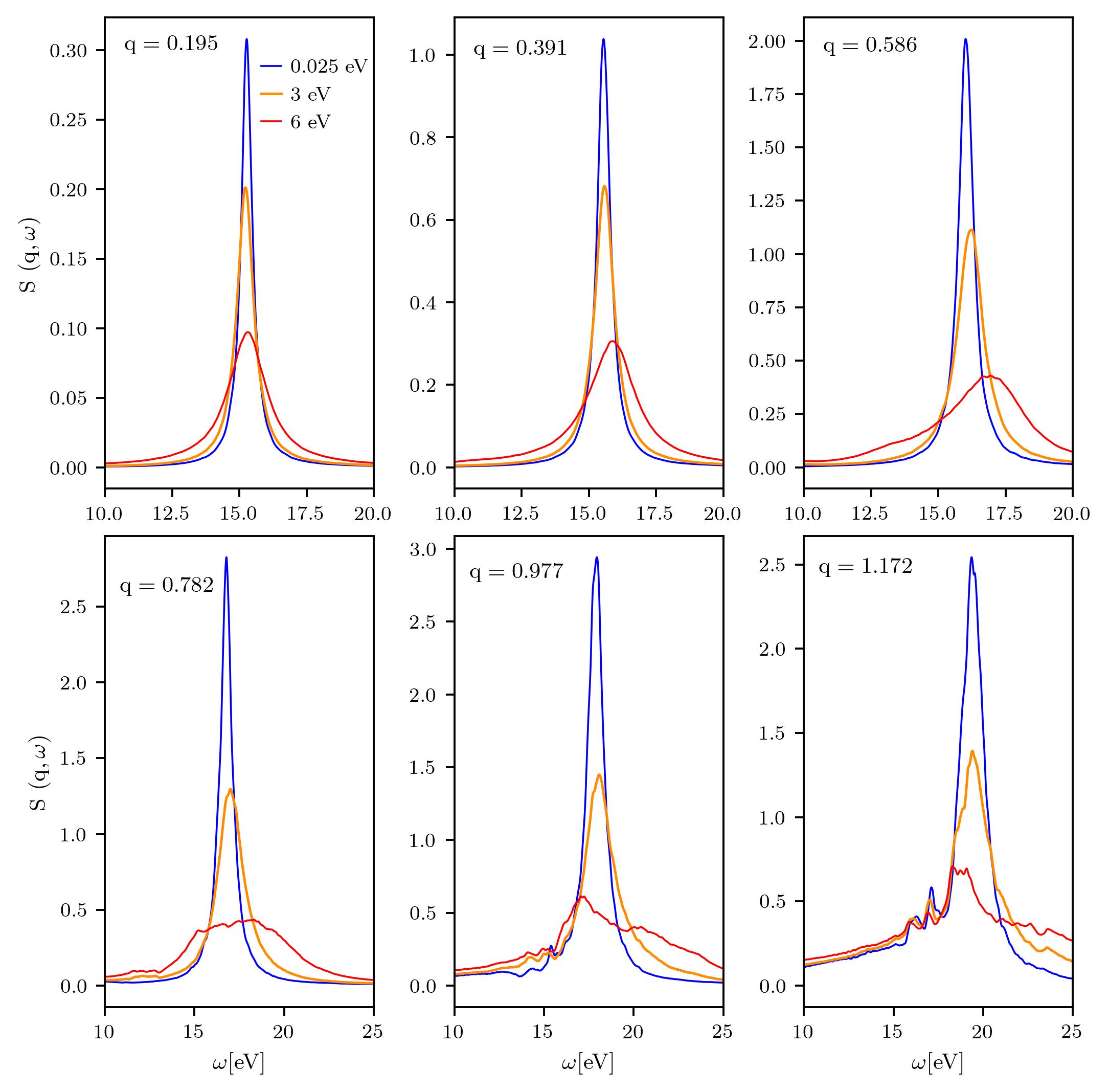}
    \caption{Results for fcc Al along the [001] direction for the ground state with $T=0.025~{\rm eV}$, and heated electrons with $T=3~{\rm eV}$ and $T=6~{\rm eV}$. The wavenumber values are given in the units of ${\rm \AA^{-1}}$.} 
    \label{fig:Al}
\end{figure*}

\section{Results} \label{s:results}
\subsection{Aluminium}

We start our discussion by analysing the results for Al. In Fig. \ref{fig:Al}, we show the DSF at $T=0.025~{\rm eV}$, $T=3~{\rm eV}$, and $T=6~{\rm eV}$ for $\vec q$ in the direction [001] and six different absolute values in the range from $0.195~{\rm \AA^{-1}}$ up to $1.172~{\rm \AA^{-1}}$.
In the quantum many-body theory, the relative change in the width of the DSF (e.g., at half maximum) with increasing wavenumber or temperature is often used as a characteristic of the change in the lifetime of the plasmon quasiparticle~\cite{mahan1990many}. 
In the ground state (ambient conditions), a plasmon is well-defined up to the plasmon cutoff wavenumber $q_c(T=0)\simeq 1.3 ~{\rm \AA^{-1}}$ \cite{Cazzaniga_2011}. At $q>q_c(T=0)$,  the plasmon enters the electron-hole continuum, where it significantly broadens and effectively decays. 
As a result, it makes sense to speak about a plasmon dispersion only at $q<q_c$, where the DSF maximum defines the plasmon energy \cite{PhysRevB.34.2097}.
From  Fig.~\ref{fig:Al}, one can see that increasing the temperature makes the DSF broader and leads to a decrease in its magnitude. Therefore, thermal excitations contribute to the plasmon broadening (i.e. dissipation) at all wavenumbers.
At $T=6~{\rm eV}$, the plasmon peak becomes ill-defined already at $q\simeq 0.8 ~{\rm \AA^{-1}}$ (i.e., we have $q_c(T)\simeq 0.8 ~{\rm \AA^{-1}}$). Accordingly, we conclude that the plasmon cutoff wavenumber reduces when the temperature is increased, which is the result of the Landau damping, i.e., the loss of energy of collective motion to the excitation of individual particles \cite{Lan67, PhysRev.164.380, bonitz_book}.

We note that the dimensionless scattering parameter defined as $\alpha=1/(q\lambda_s)$ is used to characterize the XRTS scattering regime in plasmas \cite{Glenzer_revmodphys_2009} (with $\lambda_s$ being the characteristic screening length of a test particle in plasmas \cite{moldabekov_pop_15, moldabekov_cpp_22}). For completeness, we note that  for $\alpha=1$ we have the wavenumber values $q_{\alpha=1}=2.1~{\rm \AA^{-1}}$ at $T=0.025~{\rm eV}$,  $q_{\alpha=1}\simeq 2~{\rm \AA^{-1}}$ at $T=3~{\rm eV}$, and $q_{\alpha=1}\simeq 1.88~{\rm \AA^{-1}}$ at $T=6~{\rm eV}$, where $\lambda_s$ is evaluated using the Thomas-Fermi screening length. Therefore, at considered wavenumbers we have $q<q_{\rm \alpha=1}$, i.e., a collective scattering regime.       

\begin{figure}
    \centering
    \includegraphics[width=0.5\textwidth,keepaspectratio]{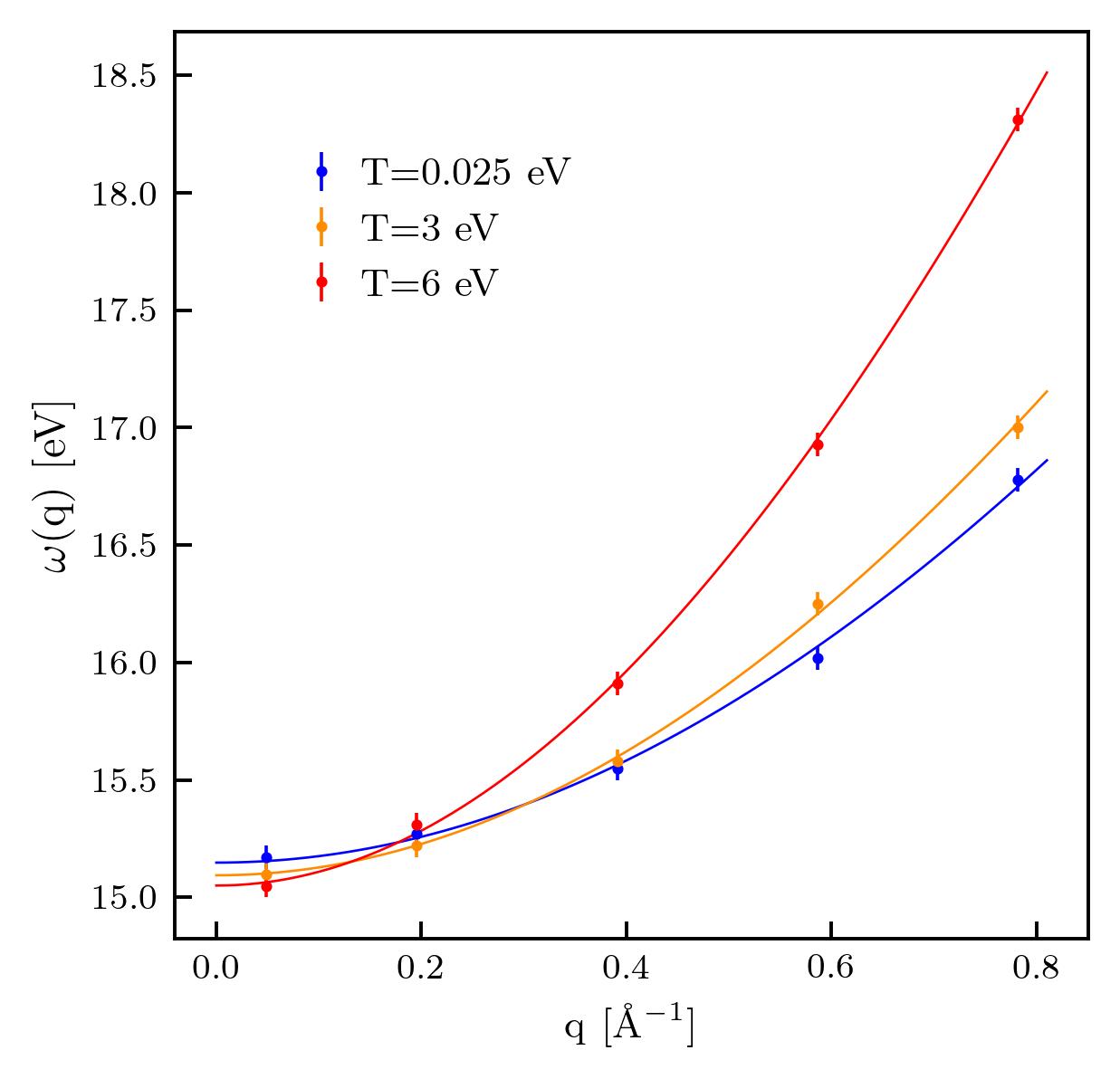}
    \caption{Plasmon dispersion along the direction  [001] in fcc Al at different temperatures. Solid lines are fit with
Bohm-Gross type quadratic dispersion. The error bars depict the uncertainty $\pm 0.05 {\rm eV}$ introduced by the Lorentzian smearing.}
    \label{fig:Al_plasmon}
\end{figure}

Next, we clearly see that the increase in temperature leads to a shift of the DSF maximum position to higher energies for $0.391~{\rm \AA^{-1}}\leq q <0.8 ~{\rm \AA^{-1}}$.
This is expected behavior similar to that of the UEG~\cite{Hamann_cpp}. However, at $q=0.195~{\rm \AA^{-1}}$, we notice that the maximum of the DSF at $T=3~{\rm eV}$ is located at slightly smaller energies compared to the DSF maximum at $T=0.025~{\rm eV}$. Although subtle, this is a rather surprising effect, which does not emerge in the UEG model where the plasmon energy in the limit $q\to0$ is independent of temperature. 
To get a better picture of this heating-induced red shift in the plasmon energy, we plot the plasmon dispersion in Fig.~\ref{fig:Al_plasmon} for $0.05~{\rm \AA^{-1}}\leq q \leq 0.782 ~{\rm \AA^{-1}}$ at $T=0.025~{\rm eV}$, $T=3~{\rm eV}$, and $T=6~{\rm eV}$. We use the Bohm-Gross type dispersion relation to fit the results from LR-TDDFT simulations for the plasmon  energy \cite{Bohm_Gross, Hamann_cpp},
\begin{equation}
    \omega^2(q)=\omega_p^2+\alpha q^2, 
    \label{eq:BohmGross}
\end{equation}
where $\omega_p$ is the plasma frequency defined in the limit of $q\to 0$, and $\alpha$ is a fitting parameter.

Within the LR-TDDFT calculations, the use of the Lorentzian smearing parameter $\eta$ leads to a broadening of the DSF introducing an uncertainty of the order of $\eta$ in plasmon the dispersion \cite{TIMROV2015460}. 
In Fig.~\ref{fig:Al_plasmon}, we depict error bars for the plasmon positions as $\pm \eta$. One can see that the plasmon dispersion has the Bohm-Gross type form at all considered temperatures. At $T=0.025~{\rm eV}$, the plasmon dispersion of Al is in agreement both with previous LR-TDDFT results and experimental data \cite{Quong_PRL}.
Increasing the temperature to $3~{\rm eV}$ and further to $6~{\rm eV}$, we observe that the plasmon energy at $q\to0$ becomes slightly reduced.

\begin{figure}
    \centering
    \includegraphics[width=0.5\textwidth,keepaspectratio]{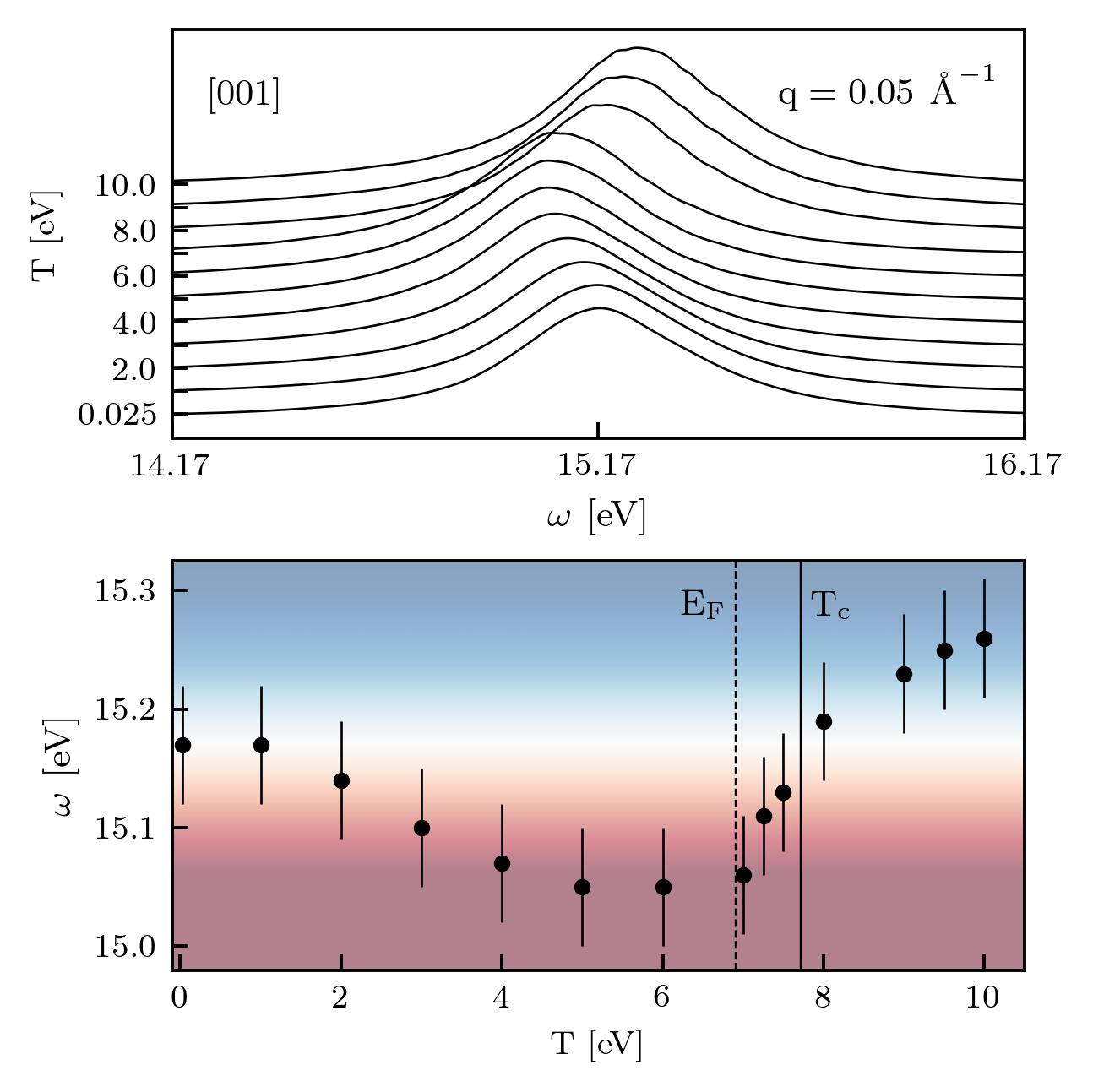}
    \caption{Top panel: the results for the DSF of electrons in fcc Al along the [001] direction.
    Bottom panel: the dependence of the plasmon energy on temperature. The results are presented for $q=0.05~{\rm \AA^{-1}}$ as $0.025~{\rm eV}\leq T \leq 10 {\rm eV}$.
    In the bottom panel, the uncertainty $\pm 0.05 {\rm eV}$ due to the Lorentzian smearing is depicted by vertical lines. The colormap in the bottom panel provides a visual representation of a blue and red shift in the plasmon energy compared to the ground state.}
    \label{fig:PlasmonSpectra}
\end{figure}

To get more information about the temperature dependence of the plasmon energy at small wavenumbers, we show the DSF (top panel) and the plasmon energy (bottom panel) at $q=0.05~{\rm \AA^{-1}}$ for fourteen different temperatures in the range $0.025~{\rm eV}\leq T \leq 10 {\rm eV}$ in  Fig. \ref{fig:PlasmonSpectra}.
In the top panel, the DSF curves are rescaled for better visualization of the change in the maximum position to have the same width at half maximum keeping the maximum position unaltered. In the bottom panel, we show the dependence of the plasmon energy at $q=0.05~{\rm \AA^{-1}}$  on the temperature and provide a quantification of the uncertainty due to the Lorentzian smearing. The vertical dashed line at $E_F=6.9~{\rm eV}$ indicates the Fermi energy (level) from the KS-DFT calculations.  
At $T<E_F$, we see that the plasmon energy at $q=0.05~{\rm \AA^{-1}}$ indeed reduces with increasing $T$.
The largest value of the red shift in the plasmon energy due to thermal excitations is close to $\delta \omega_p\simeq 0.1~{\rm eV}$. 
At $T\gtrsim E_F$, the plasmon energy starts to increase with $T$. The vertical solid line at $T_c=7.72~{\rm eV}$ specifies where the chemical potential of the valence electrons becomes negative (i.e., at $T>T_c$). For comparison, the chemical potential of the UEG at the same density becomes negative at $T\gtrsim 13.75 ~{\rm eV}$.
Finally, the finite-temperature plasmon energy $\omega_p(T)$ becomes larger than the ground-sate plasmon energy $\omega_p(T=0)$ at $T\gtrsim T_c$.
We note that at all considered temperatures up to $10~{\rm eV}$, the plasmon energy remains smaller than the plasmon energy computed using the UEG model ($\omega_p=16~{\rm eV}$).

\begin{figure}
    \centering
    \includegraphics[width=0.5\textwidth,keepaspectratio]{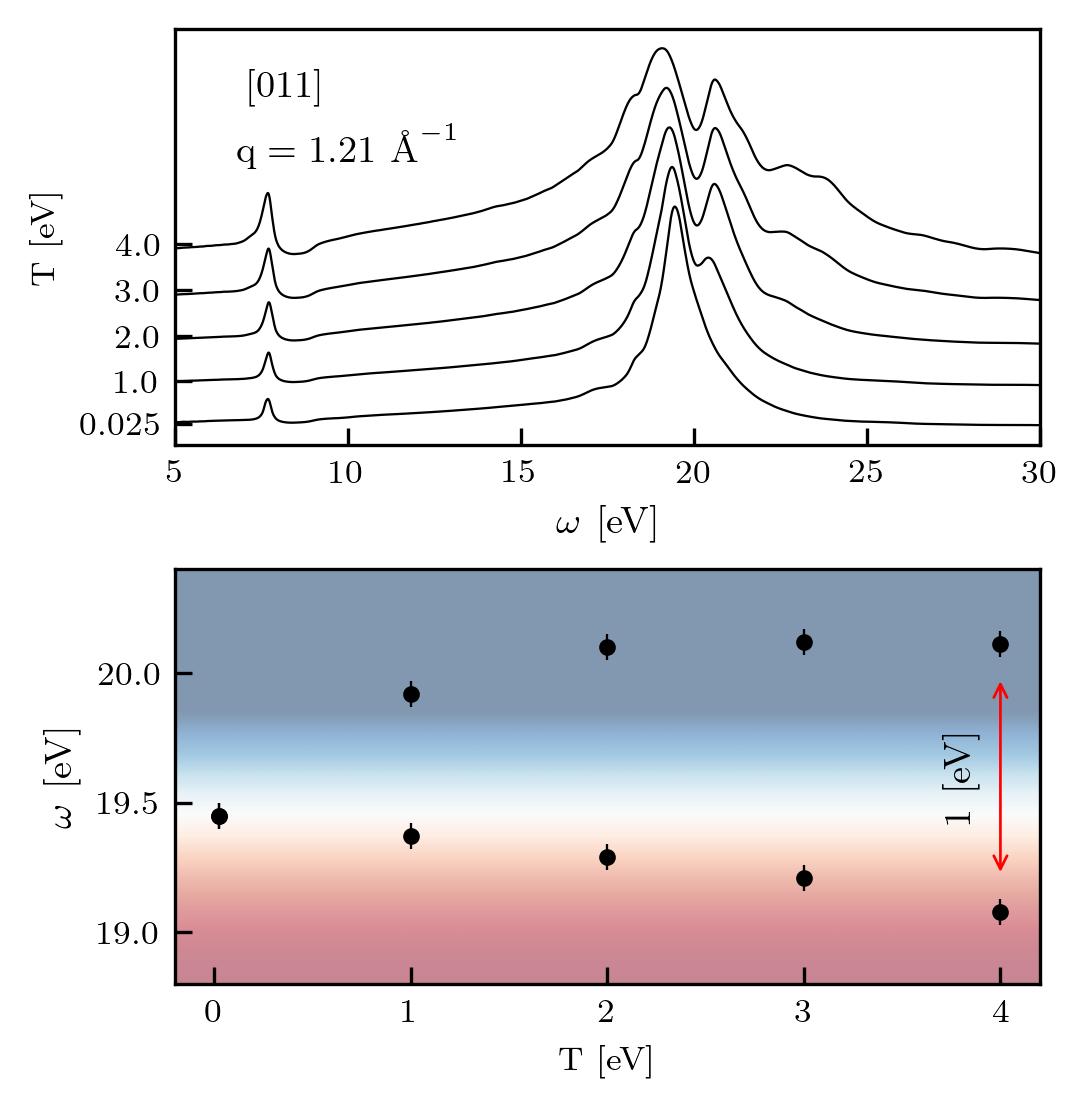}
    \caption{Top panel: the DSF of electrons in fcc Al along the [011] direction.
    Bottom panel: the dependence of the plasmon and the double-plasmon excitation energy on temperature, where the vertical lines indicate the uncertainty $\pm 0.05 {\rm eV}$ in the maximum location due to the Lorentzian smearing. The results are presented for $q=1.21~{\rm \AA^{-1}}$ at $0.025~{\rm eV}\leq T \leq 4 {\rm eV}$. The colormap in the bottom panel provides a visual representation of a blue and red shift of the peaks in the double-plasmon feature compared to the ground state. }
    \label{fig:Al_011}
\end{figure}

The DSF of electrons in fcc Al is nearly isotropic at small wavenumbers $q<1~{\rm \AA^{-1}}$ and becomes anisotropic at large wavenumbers $q\gtrsim 1~{\rm \AA^{-1}}$ \cite{PhysRevB.40.5799}.
As an example of the ion-lattice induced anisotropy, we show the DSF for the [011] direction at $q=1.21~{\rm \AA^{-1}}$ for  $0.025~{\rm eV}\leq T \leq 4~{\rm eV}$ in Fig. \ref{fig:Al_011}.
At $T=0.025~{\rm eV}$, we have a well-defined plasmon peak since $q<q_c$. In addition, we see a smaller well-visible peak at $\omega \simeq  7.7 {\rm eV}$.
This structure was observed in experiments \cite{Larson_2000} and was shown to be entirely due to lattice effects in previous LR-TDDFT based studies \cite{Cazzaniga_2011}.
From Fig.~\ref{fig:Al_011}, we further see that the peak at $\omega \simeq  7.7 {\rm eV}$ is only weakly affected by thermal excitations at  $T \leq 4~{\rm eV}$.
A much more interesting result of thermal effects is the emergence of a double-plasmon excitation with the increase in temperature.
The double-plasmon excitation emerges already at $T=1~{\rm eV}$ and becomes more distinct up to $T=3~{\rm eV}$.
At $T=4~{\rm eV}$, we observe that the DSF further splits into more local peaks. The onset of such splitting into multiple peaks is already possible to detect at  $T=2~{\rm eV}$. 
A similar effect has been observed experimentally in the ground state as well as using LR-TDDFT simulations at $q>q_c$ \cite{Schuelke, Larson_2000, Cazzaniga_2011}, where it was interpreted to be a band-structure effect.
It is known that the adiabatic approximation for the XC-kernel tends to over-amplify the double-plasmon excitation and the use of dynamic kernels somewhat smooths out the split but is unable to wash out it entirely \cite{Cazzaniga_2011}.  In the ground state,  the splitting of the plasmon peak to form the double-plasmon excitation starts to take place with increasing wavenumber after entering the electron-hole continuum at $q>q_c$, i.e., where the plasmon becomes strongly damped. 
As we discussed earlier and showed in Fig. \ref{fig:Al}, the increase in temperature moves the plasmon-damping region to the smaller wavenumbers. 
Therefore, the increase in temperature at fixed wavenumber pushes the DSF closer to the plasmon-damping region resulting in the emergence of the double-plasmon excitation. 
In the bottom panel of Fig. \ref{fig:Al_011}, we show the dependence of the plasmon and the double-plasmon peak positions on temperature.
From this figure, we see that at $1 ~{\rm eV} \leq T\leq 4~{\rm eV}$, the segregation of peaks of the double-plasmon is the order of $1~{\rm eV}$.
This is within the range of the experimental accuracy at the European XFEL as we discuss it is Sec. \ref{s:exp}. 

\begin{figure}
    \centering
    \includegraphics[width=0.5\textwidth,keepaspectratio]{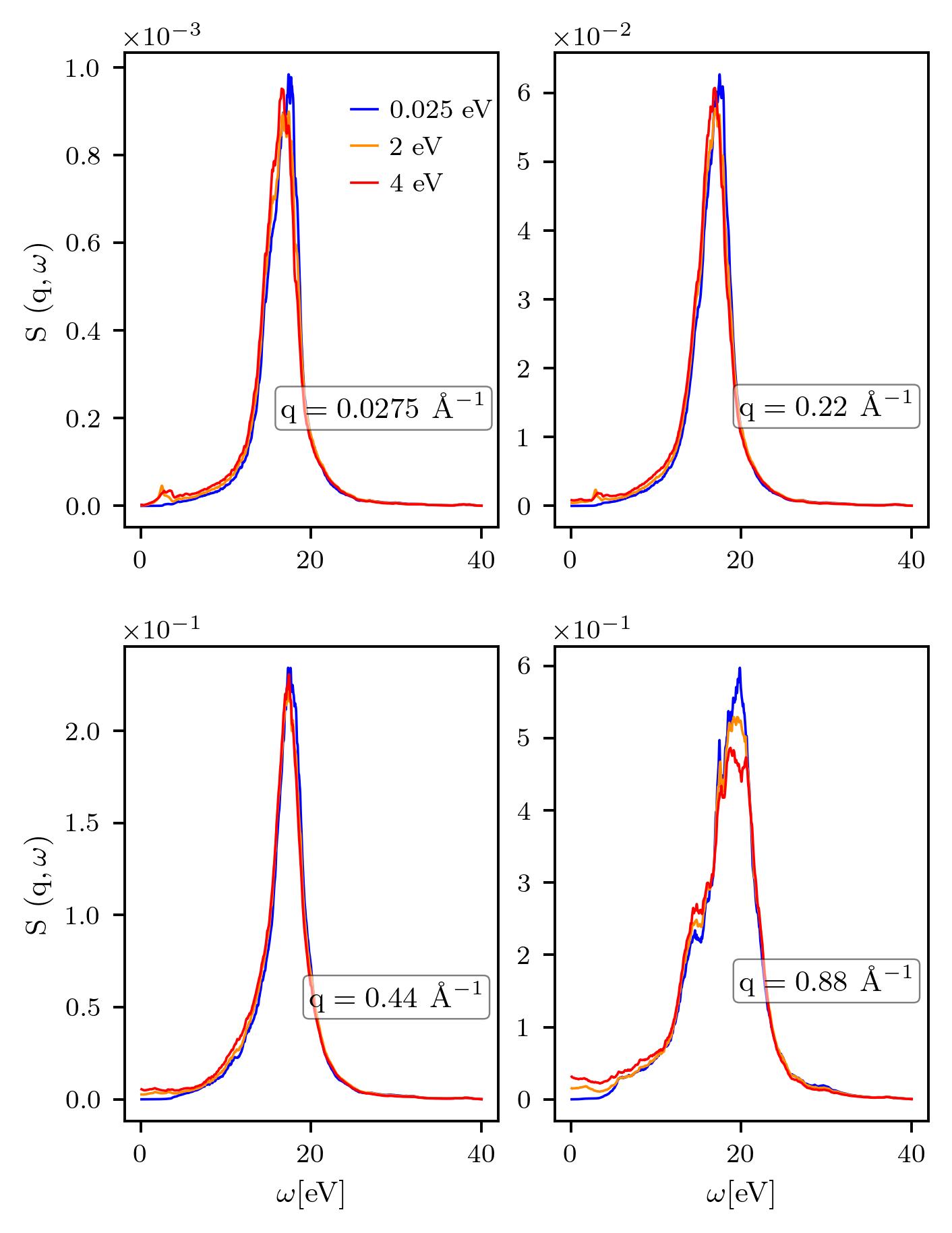}
    \caption{Results for Si along the [001]  direction for the ground state with $T=0.025~{\rm eV}$, and heated electrons with $T=2~{\rm eV}$ and $T=4~{\rm eV}$. 
    }
    \label{fig:Si_001}
\end{figure}

\begin{figure}
    \centering
    \includegraphics[width=0.5\textwidth,keepaspectratio]{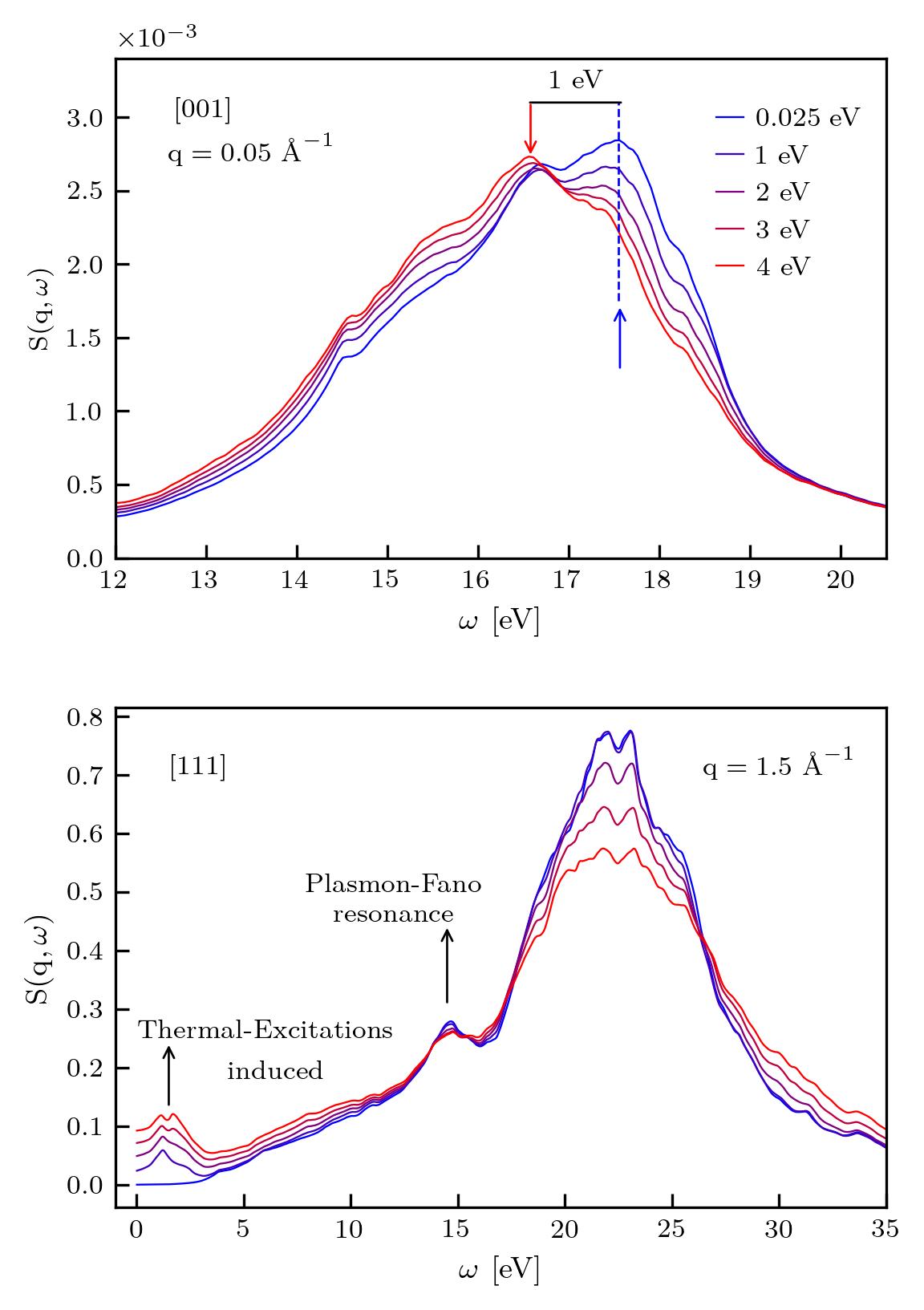}
    \caption{Results for Si along the [001] direction at $q=0.05~{\rm \AA^{-1}}$ (top panel) and the [111] direction at $q=1.5~{\rm \AA^{-1}}$ (bottom panel).
    The results are presented for  $0.025~{\rm eV}\leq T \leq 4 {\rm eV}$. Here we use colors to differentiate between varying temperatures.
    }
    \label{fig:Si_111}
\end{figure}


\subsection{Silicon}
To see whether the above-presented results for the thermal excitations in Al can occur in semiconductors, next we consider Si.
We note that laser irritation-induced heating of electrons can lead to lattice instability due to the weakening of the silicon bond \cite{PhysRevB.26.1980, PhysRevLett.96.055503}.
Therefore, we consider relatively low electronic temperatures $T\leq 4~{\rm eV}$. 
First, we show in Fig. \ref{fig:Si_001} the DSF of the electrons in Si in the direction [001] at $T=0.025~{\rm eV}$, $T=2~{\rm eV}$, and $T=4~{\rm eV}$ for wavenumbers in the range $0.0275~{\rm \AA^{-1}}\leq q \leq 0.88~{\rm \AA^{-1}}$ ($q<q_c\simeq 1.14 ~{\rm \AA^{-1}}$); it is less sensitive to temperature changes compared to that in Al.
Indeed, the increase in the temperature to $4~{\rm eV}$ almost has no effect on the DSF width (e.g., at half maximum). In contrast, we observed significant broadening already at $3~{\rm eV}$ for Al.
In this regard, we note that the DSF maximum in Si is usually referred to as a plasmon, but does not follow the Bohm-Gross type dispersion at small wavenumbers \cite{Schuelke_Si}. For completeness, we mention that our LDA-based calculations yielded the Fermi energy $5.37~{\rm eV}$  and a band gap of $0.5~{\rm eV}$ at ambient conditions.
We note that the LDA-based calculations underestimate the band gap value about two times. Nevertheless, this is not problematic for the DSF at finite wavenumbers, where the LR-TDDFT based on the LDA provides an accurate description of the experimental XRTS signal at a finite momentum transfer \cite{Si_Weissker}. This is in contrast to the properties in the optical limit, e.g., for the absorption spectra \cite{Onida_revmodphys_2002}.

From Fig. \ref{fig:Si_001}, we notice two interesting features that appear with the increase in the temperature: (i) a slight red shift of the plasmon energy at $0.0275~{\rm \AA^{-1}}$  and (ii) the increase of the DSF values at small frequencies (particularly noticeable at $q= 0.88~{\rm \AA^{-1}}$). From Fig. \ref{fig:Si_001}, we see that the red shift of the plasmon energy diminishes with the increase in the wavenumber. Similar behavior we observed for Al (see Fig. \ref{fig:Al_plasmon}). 

To gain more insight on the former effect, we show results for the DSF for $q=0.05~{\rm \AA^{-1}}$  at five different temperatures in the range  $0.025~{\rm eV}\leq T \leq 4~{\rm eV}$ in the top panel of Fig. \ref{fig:Si_111}. We can clearly see the red shift of the plasmon energy due to thermal excitations. 
 For Si, the red shift reaches $\delta \omega_p\simeq 1~{\rm eV}$, which is much larger than the uncertainty $\pm 0.1 {\rm eV}$ introduced by the Lorentzian smearing.
 Also, we note that the plasmon red shift in Si ($\delta \omega_p\simeq 1~{\rm eV}$) is more significant than in Al ($\delta \omega_p\simeq 0.1~{\rm eV}$).
 In addition, the results for Si and Al indicate that the red shift of the plasmon energy due to thermal excitations is a generic effect for both semiconductors and metals.

Finally, we consider the direction [111], which was extensively studied at ambient conditions experimentally and theoretically \cite{Schuelke_Si, Si_Weissker, Weissker_PRL}.  In the bottom panel of Fig. \ref{fig:Si_111}, we provide results for the DSF in the direction [111] at $q=1.5~{\rm \AA^{-1}}$.
The main broad peak of the DSF in the bottom panel of Fig. \ref{fig:Si_111} is usually referred to as a plasmon, although it has some fine structure due to band-structure effects at $q>q_c$.
The second peak to the left of the plasmon is the so-called plasmon-Fano resonance at around $15~{\rm eV}$ originating from local-field effects \cite{Schuelke_Si, Si_Weissker, Weissker_PRL}.
The thermal excitations broaden the width of the plasmon and the plasmon-Fano resonance but do not have a noteworthy effect on their positions.
At low energies $\omega<5 ~{\rm eV}$, one can clearly see a signature induced by thermal excitations. 
This increase in the DSF at low energies due to thermal effects at $T=1 {\rm eV}$ ($T=4 {\rm eV}$) reaches approximately $10\%$ ($20\%$)  of the magnitude of the main plasmon peak. This is in contrast to the DSF in the  [001] direction shown in Fig. \ref{fig:Si_001}, where these thermal features are rather weak and do not exceed $5\%$ of the magnitude of the plasmon peak. Physically, these features might originate from transitions between accumulated states located closely above the Fermi level since heating results in partial occupation of these levels, allowing new excitation features to emerge. Recently, we reported a similar effect for copper \cite{moldabekov2024ultrafast}.

As we discuss in Sec. \ref{s:exp}, the thermally-induced features at low frequencies in the [111] direction can be observed in experiments, e.g., at European XFEL, where a beam with a width of $\Omega<1~{\rm eV}$ can be produced for XRTS diagnostics~\cite{Descamps2020, McBride_RSI, Wollenweber_RSI}.

\section{The possibility of detecting thermal excitation induced signals at a few electronvolts}\label{s:exp}

A system of hot electrons in a crystal structure of ions can be produced, e.g., using an FEL pulse with a duration of $\sim 10-100$ fs.
In the case of self-scattering experiments, scattering from a range of electron temperatures takes place. We therefore expect that the plasmon red shift in Al will not be directly observable in self-scattering experiments due to how small of a shift it is.  However, the plasmon red shift in Si is much larger (about $\sim 1$~eV), and so the overall change in the plasmon shape may be more readily observed. 
Though the spectrum is time- (and temperature-) integrated, it is anticipated that, for a sufficiently narrow beam, the second peak in the double-plasmon signature (top panel of Fig.~\ref{fig:Al_011}) may be observable given that it is fairly distinguishable from the primary plasmon position. Similarly, the thermally-induced small-$\omega$ excitation in Si (see Fig.~\ref{fig:Si_111}) should develop as an additional bump near or on the elastic peak.  
The aforementioned issues of self-scattering experiments can be entirely avoided by driving electrons to high temperatures using a 400 nm optical short-pulse laser on thin targets, and tuning the delay of the FEL probe, as was reported in Refs. \cite{Mo_Science_2018, Chen_prl_2013}. 

Second, probing the thermal signatures of electrons in an ionic crystal structure requires a sufficiently narrow beam.
We anticipate that the self-seeding technique creating a seeded beam with a FWHM of $\Omega \lesssim1$~eV \cite{Zastrau_2021} could allow one to detect the double-plasmon signature in Al and the plasmon red-shift as well as the small-$\omega$ thermal excitations in Si.
Additionally, we note that to further improve the probe quality, a monochromator can be placed in the beam to cut out the SASE pedestal while retaining the seeded spike. 

Third, besides of the beam quality, an equally important aspect of the XRTS diagnostics is the detector (spectrometer) resolution. The signals reported in this work can be measured by using high-resolution diced crystal analyzers with resolutions $\delta_{\rm \omega}\simeq 40-50$~meV fielded at FEL facilities~\cite{Wollenweber_RSI,Descamps2020,McBride_RSI}.
Indeed, this was recently demonstrated for the XRTS measurement with ultrahigh resolution of the plasmon dispersion in Al at ambient conditions  \cite{gawne_prl}.

We conclude that the current capabilities at the European XFEL provide a possibility to detect the plasmon red shift in Si, the low energy excitations in the XRTS of Si, and the double-plasmon signature in Al.  The plasmon red shift in Al could be inferred from the plasmon dispersion at different $q$ vectors. Indeed, as shown in Fig.~\ref{fig:Al_plasmon}, the plasmon dispersion is well-described by the Bohm-Gross dispersion relation, Eq.~\ref{eq:BohmGross}.
Therefore, by making measurements at several different scattering angles and fitting the Bohm-Gross relationship to the measured dispersion, the plasma frequency at $q=0$ can be inferred.

\section{Conclusions}\label{s:end}

We have performed a systematic investigation of the effect of thermal excitations on the DSF of electrons in Al and Si based on LR-TDDFT simulations within the ALDA.
The results show that there is a wealth of interesting excitation signatures that can be observed at XFEL facilities such as the European XFEL. 
For Al, LR-TDDFT predicts a thermal excitation induced red shift of the plasmon energy up to $\delta \omega\simeq 0.1~{\rm eV}$ at $T\lesssim 5 {\rm eV}$. 
This red shift is also predicted for Si at $T\lesssim 4 {\rm eV}$, with the largest value reaching $\delta \omega\simeq 1~{\rm eV}$.
Furthermore, the simulation results show that an increase in the temperature leads to the formation of a double-plasmon peak in Al, which is caused by the shift of the Landau damping region to smaller wavenumbers. Finally, we observe the emergence of a local peak in the DSF of Si at low energies $\omega<5 ~{\rm eV}$, with a magnitude of between about $10\%$ and $20\%$  of the plasmon peak at $1\leq T\leq 4~{\rm eV}$.

We provided a discussion of the possibility of measuring these effects in both considered elements.
For Si, the plasmon red shift and the increase in the DSF value at small frequencies are within the range of the experimental capabilities at the European XFEL.
Similarly, the formation of the double-plasmon peak due to thermal effects in Al should be measurable as well.
The direct observation of the $0.1~{\rm eV}$  plasmon red shift in Al at small wavenumbers will be problematic.
However, by fitting the Bohm-Gross dispersion relation at several $q$ values with temperature might allow one to measure the onset of the plasmon red shift.

The results presented here can be refined using more complicated approximations for the XC kernel, if the results of experiments on isochorically heated electrons will render the ALDA approximation used here insufficiently accurate. In this way, the controlled set-up and highly accurate diagnostics at modern XFEL facilities constitute an important opportunity to rigorously benchmark dynamic simulation methods for WDM research.

We believe that there are many more interesting phenomena to be explored by XRTS experiments in X-ray driven materials.
In  Al and Si that have been considered in this work,  the electronic coupling parameter is about $r_s\simeq 2$, with $r_s\sim n^{-1/3}$ being defined by the ratio of the Wigner–Seitz radius to the first Bohr radius.
The electronic structure is known to have more features in the DSF at lower densities due to stronger inter-particle correlations, e.g., in alkali metals.
For instance, in cesium electrons have a negative plasmon dispersion and form a so-called roton minimum \cite{Alkali_Metals, PhysRevB.40.10181}. 
Another example are lithium-based materials, such as lithium hydride, where electrons are strongly correlated with $r_s\sim 10$ \cite{Lithium_Ammonia}. 
Therefore, the study of quantum-many-body effects arising from the interplay between thermal excitations and inter-particle correlations in isochorically heated electrons in an ion crystal lattice remains a largely unexplored promising research field for future studies.

On a final note, we observe that an ideal experiment would employ distinct X-ray pump and X-ray probe beams that are separated in colour and time. Good colour separation would allow one to measure the scattering from the probe beam while avoiding collecting the scattering signal from the pump beam. Meanwhile, good time separation between the beams would allow heating to be complete by the pump before the probe is incident. We would therefore avoid the aforementioned limitations of self-scattering experiments.
Furthermore, varying the time separation between the two pulses would allow investigation into the establishment of electron thermal equilibrium in femtosecond X-ray experiments.
Such X-ray pump X-ray probe capabilities have been demonstrated at existing XFEL facilities, such as the at European XFEL~\cite{Liu2023} and at SACLA in Japan~\cite{hara2013two,inoue2020two}, through the use of an electron chicane. In this setup, the pump and probe beams travel along the same beamline, and therefore monochromation -- which is necessary to observe the thermal excitation signals reported in this work -- would not be currently possible. Additional care would also be needed in the choice of photon energies for the pump and probe in order to still be able to focus the beams using existing beryllium compound refractive lenses. Nevertheless, we anticipate that the further development of X-ray pump X-ray probe systems with high resolution would open up new avenues for exploring the details of nonequilibrium transient states.


\begin{acknowledgments}
This work was partially supported by the Center for Advanced Systems Understanding (CASUS), financed by Germany’s Federal Ministry of Education and Research (BMBF) and the Saxon state government out of the State budget approved by the Saxon State Parliament. This work has received funding from the European Research Council (ERC) under the European Union’s Horizon 2022 research and innovation programme
(Grant agreement No. 101076233, "PREXTREME"). 
Views and opinions expressed are however those of the authors only and do not necessarily reflect those of the European Union or the European Research Council Executive Agency. Neither the European Union nor the granting authority can be held responsible for them. Computations were performed on a Bull Cluster at the Center for Information Services and High-Performance Computing (ZIH) at Technische Universit\"at Dresden, at the Norddeutscher Verbund f\"ur Hoch- und H\"ochstleistungsrechnen (HLRN) under grant mvp00024, and on the HoreKa supercomputer funded by the Ministry of Science, Research and the Arts Baden-W\"urttemberg and
by the Federal Ministry of Education and Research.
\end{acknowledgments}

\bibliography{bibliography}
\end{document}